# On measuring team stability in cooperative learning: An example of consecutive course projects on software engineering

Yanqing Wang*, Hong Ge, Xiaojing Feng, Jie Yu
School of Management, Harbin Institute of Technology, Harbin 150001, China

**Abstract:** Cooperative learning theory has shown that stable membership is a hallmark of effective work teams. According to relation strength and social network centrality, this paper proposes an approach to measure team stability reasons in consecutive cooperative learning. Taking consecutive course projects of software engineering in a university as examples, we examine the relation between team stability and learning performance in consecutive cooperative learning from two parts: learning score and learning satisfaction. Through empirical analysis, it arrives at the conclusion that learning score is in weak positive correlation with team stability. Through questionnaire and interviews, it finds out 78% of the students did not value the importance of team stability, and 67% of the teachers never recommend the students to keep stable teams. Finally, we put forward an expected correlation model of learning performance as future work and discuss instability as well.

**Keywords:** consecutive cooperative learning; team stability; social network; course project; learning performance

## 1. Introduction

Cooperative learning captures increasing attention of the scholars from diverse fields. Slavin (1995) proposed that cooperative learning is a series of learning activities of the students in teams and a teaching skill to encourage or accept them in line with the whole team's performance. Also, cooperative learning refers to all relevant actions of students achieving the goal of joint learning in teams for learning performance with the help of cooperation and some incentive mechanisms (Quarstein, 2001; Liu & Huang, 2002). May and Doob (1937) found that people who cooperate and work together to achieve shared goals, were more successful in gaining outcomes than those who strived independently to complete the same goals.

In the context of cooperative learning, team stability is highly concerned (Carley, 1991; Johnson & Johnson, 1994). In the Oxford Dictionary of Sports Science & Medicine, team stability is explained as the degree to which the membership of a team remains the same. It can be defined in the dimension of time team members remain together. Moreland, Argote, and Krishnan (1998) showed that stable team membership simplifies learning and intra team coordination. Savelsbergh et al (2010) confirmed a strong direct relation between team stability and team learning. Edmondson et al. (2007) in their review on team learning stated that teams with a more stable composition provide higher rates of improvement. Especially when it comes to learning by doing they claim team stability to be an influencing reason. The extent to which members have worked together is clearly an important issue for understanding how well they share their knowledge, skills, and actions to achieve collective aims. On the other hand, teams characterized by a lack of group longevity experience greater difficulty recognizing and integrating their knowledge for efficient task completion (Liang, Moreland, & Argote, 1995).

Similarly, project team stability in product development draws much attention. In a new product development context, project team stability refers to the extent to which the core members of a cross-functional team remain during the whole project, from project approval to product launch (Slotegraaf & Atuahene-Gima, 2011). From a knowledge-based perspective, stability stimulates collaboration (Pelled, Eisenhardt, & Xin, 1999), thus enabling a cross-functional new product development team to take apart the knowledge barriers that hinder innovation success (Carlile, 2002). Slotegraaf and Atuahene-Gima (2011) argued that stability in a new product development team may offer important advantages for decision making. Yeh et al (2005) found an R&D team's longevity and stability is positively relevant to its performance.

The research on cooperative learning is a road full of hardships. Sharan (2010) regarded the constant evolution of cooperative learning as a threat. Teachers may become confused and lack an understanding of it. Especially, few research on team's stability in consecutive cooperative learning has been found. As for major issues such as random teaming, weak consistency of most project teams in undertaking consecutive course projects of software engineering, Wang et al (2008) put forward the strategic framework featuring "stable teaming and consistent topics". But their research was at a preliminary level without bringing forth any feasible measuring methods. This infers the relation between team stability and learning performance, especially in consecutive cooperative learning, may be more complex than the literature has pointed out.

To find out these potentially complex relations, we explore the measuring methods of team stability in consecutive cooperative learning. Learning performance has been successfully measured from learning score and learning satisfaction (Sun & Cheng, 2007; Shaw, 2010). Therefore, taking the consecutive course projects of a School of Software at Harbin Institute of Technology(SoS@HIT) in China as examples, we discuss the team stability's impact on learning performance from the following aspects: we analyse the team stability's influence on learning score from empirical investigation, and the team stability's influence on learning satisfaction from the questionnaires and the interview. Last but not the least, we put forward an expected model of learning performance and discuss team instability preliminarily.

## 2. Team stability in consecutive cooperative learning

While team stability is defined by time the team members remain together in the Oxford Dictionary of Sports Science & Medicine, the definition is so simple that we cannot apply it in consecutive cooperative learning.

### *2.1 Relation strength*

Cooperative learning between the team members forms a set of cooperative learning network which is in social network (Gupta, 2004). The following adopts the method of social network analysis to conduct quantitative analysis of team stability in consecutive cooperative learning. There into, the relation between the team members is one of the core research objects in consecutive cooperative learning. The relation contains three characteristics: the content, the direction and the strength (Qu, 2008). In consecutive cooperative learning, the content of relation refers to that of cooperative learning of the two relational participators; the direction of relation is temporarily null; the strength of relation has undergone the following discussion and definition: it is "a mixture of mutual helpful service featured by quantity of time, strength of emotions, intimacy (shared trust) and connection". In realities, the four indexes namely the persistent period, frequency, intimacy and the topic selection scope of course projects in this research set up the strength of the relation. Table 1 below summarises some extensive and potential relation strength and the matching indexes of cooperative learning (Wang, 2009).

*Table 1. Some components of relation strength*

| Index of relation strength | Matching index of cooperative learning |
|---|---|
| Frequency | Times of cooperation |
| Intimacy/Closeness | Relationship between students |
| Requirement to partners | Teaming strategy |
| Diversity of association in society | Topic selection of course project |
| Persistent period | Duration of a cooperation |
| Strength to provide support | Feeling |
| Ability of social communication | Disposition of students |

Findings through summarizing above relation strength indexes of the practical social network show the measuring method of relation strength has not been proposed well. To quantify such relation strength in consecutive cooperative learning is specifically to describe and reflect the relation strength of the two participators through the frequency of cooperative learning. The research of the previous scholars has provided the measuring approaches of relation strength in some network and the most typical is Salton's measuring approach of the relation strength of international partners between two countries and territories, as showed in E.1(Lv, 2009).

$$R_{ij} = c_{ij} \div \sqrt{c_i \times c_j}, i \neq j \qquad \text{E.1}$$

There into, $c_{ij}$ is the frequency of cooperation between the two territories; $c_i$ and $c_j$ respectively represent the frequency of the territories $i$ and $j$ in their own participation; $R_{ij}$ is the relation strength between participators $i$ and $j$.

The method depicts the relation between the nodes in the network and their intimacy degree between nodes, meanwhile it is also applicable to calculate the static relation strength between the nodes in the consecutive cooperative learning network when regarding the individuals as different countries, therefore the measuring approach of the total relation strength of a certain participator (E.2).

$$R(i) = \sum_{j=1}^{n} (c_{ij} \div \sqrt{c_i \times c_j})(i \neq j) \qquad \text{E.2}$$

There into, $R(i)$ is the total relation strength of participator $i$ has participated with the other $j$ participators, $n$ is the number of students who cooperate with participator $i$, so the value of $R(i)$ is the sum of $R_{ij}$ in $n$ times.

*2.2 Improved network centrality*

The method above can help calculate the relation strength of each node in cooperative learning network. However, in consecutive cooperative learning, the changes of relation strength have a certain time sequence. So it is inaccurate to quantify the frequencies of cooperative learning simplify.

In the social network, assessing partners in cooperation is to analyse the position namely the *centrality* (importance) of the cooperator in the cooperative network (Luo, 2005).

There are some familiar methods in measuring centrality involving adjacency matrix which we apply in this topic to regard the cooperative team members in consecutive course projects as the nodes in the network. If they are in partnership, the adjacency matrix will have a corresponding value of 1, otherwise, 0. Consecutive cooperation touches the idea of measurement of centrality. Represent the centrality of a node through the degree of the node in the network (Krackhardt & Hanson, 1993) and in the calculation method in E.3.

$$C(i) = \sum_{j=1}^{n} r_{ij} \qquad \text{E.3}$$

There into, $C(i)$ is the centrality of the $i^{th}$ student; $n$ is the number of the students; $r_{ij}$ is the element in the adjacency matrix. If students $i$ and $j$ have cooperated, then $r_{ij}=1$, otherwise, $r_{ij}=0$.

Here we introduce the centrality into calculating team stability and weight the centrality through adding damped exponential to get the relation of quantification of centrality between the cooperators in the dimension of time. The *centrality* of the student $i$ after improvement is proposed, as depicted in E.4.

$$C'(i) = \sum_{j=1}^{n} \sum_{q=1}^{m} (r_{ij} \times (e^{-1})^{m-q}) \qquad \text{E.4}$$

There into, $r_{ij}$ is 0 or 1. When students $i$ and $j$ have cooperated in the same team in the $q^{th}$ cooperative learning, $r_{ij}=1$, otherwise, $r_{ij}=0$; $m$ is the number of the completed cooperative learning activities; $q$ is the $q^{th}$ cooperative learning obligation of student $i$, $1 \leq q \leq m$. We apply the $e^{-1(m-q)}$ as damped exponential since we believe that the impact of two closer cooperations is much stronger than that of the cooperations which happened faraway.

*2.3 Team stability factor*

For the theories relevant to calculation of relation strength in social network and combination of the improved *centrality* proposed in E.4, it puts forward a measurement to the factors of *team stability* in consecutive cooperative learning, as depicted in E.5.

$$S(i) = \sum_{j=1}^{n} \left\{ \left[ \sum_{q=1}^{m} \left( r_{ij} \times \left( e^{-1} \right)^{m-q} \right) \right] \times R_{ij} \right\}, (i \neq j) \qquad \text{E.5}$$

There into, $S(i)$ is the team stability factor of student $i$ to complete the $m$ cooperative learning tasks; the values of $r_{ij}$ and $q$ are the same as in Formula 4; see Formula 1 for the value of $R_{ij}$. Additionally, students $i$ and $j$ are all from the set of the students and are two different people.

## 3. Investigation design

The data used in this paper came from the consecutive course projects of software engineering specialty of SoS@HIT by the students of three grades. The four course projects include data structure, database, software engineering and integrated course projects spread in the four semesters of the second and third grades, as depicted in Fig. 1.

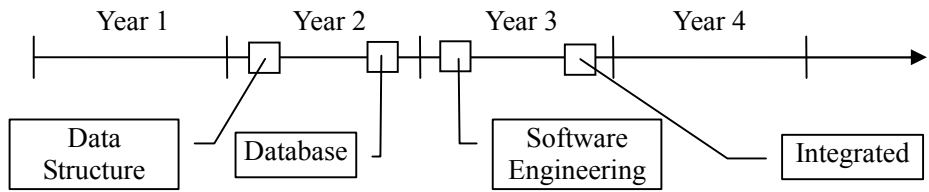

*Fig. 1. Four consecutive course projects of undergraduates in SoS@HIT*

In each course project, a project team consists of 3 to 5 students in a class and selects a topic from the topic database or the topic of their own. After two weeks of development, each team needs to put in their complete solution of the project. The teachers give scores to each team and to each individual student according to the quality of project, students' performance etc. With the help of SoS@HIT library, we collect the data relevant to the four course projects by the students who graduated from 2008 through 2010, including the *team ID*, *total score*, *the topic of each team*, *student no*, *individual score* etc. We have entered related data in the SQL Server database with the sample rows, as shown in Table 2 and Table 3.

For convenience of research, we make the following assumptions:

- The four consecutive cooperative activities occur in the same environment;
- Teachers give marks to students objectively;
- Students have not been given the idea of team stability, namely, they form teams naturally.

*Table 2. Data samples in table TeamList*

| ID | Grade | Class | Course | Score | LeaderNo | Topic |
|---|---|---|---|---|---|---|
| 3762 | 2006 | 0637101 | Data Structure | 86 | 063710120 | Traffic guidance system |
| 3763 | 2006 | 0637101 | Data Structure | 94 | 063710107 | Gobang game with GUI |
| 3764 | 2006 | 0637101 | Data Structure | 94 | 063710124 | Barbershop simulation system with Queue |
| 3765 | 2006 | 0637101 | Data Structure | 94 | 063710112 | Ambulance schedule simulation |
| 3767 | 2006 | 0637101 | Data Structure | 80 | 063710117 | Traffic guidance system |
| … | … | … | … | … | … | … |

*Table 3. Data samples in table Score*

| TeamID | StudentNo | Score |
|---|---|---|
| 3762 | 063710120 | 87 |
| 3762 | 063710127 | 86 |
| 3762 | 063710116 | 86 |
| 3762 | 063710115 | 78 |
| 3763 | 063710107 | 92 |
| 3763 | 063710113 | 91 |
| … | … | … |

## 4. Correlation analysis

### 4.1 Correlation of team stability with learning score

We calculated the *mean score, S(i)* etc of each student with the methods defined in Formula 5, and data as displayed in Table 3. We can see the results in Table 4.

*Table 4. Data samples in table FinalResults*

| StudentNo | S | MeanScore |
|---|---|---|
| 1063710323 | 2.74 | 87.50 |
| 1043710120 | 0.54 | 76.00 |
| 1043710129 | 1.48 | 89.33 |
| … | … | … |

In the correlation analysis as follows, we regard the score as the dependent variable and team stability is the independent variable. The regression analysis results got by the statistic package for social science (SPSS) are listed as in Formula 6 according to table 5.

$$Learning\ score\ =\ 82.114\ +\ 0.502 \times S \qquad (6)$$

There into, the data in Table 5 report the results of T statistic and F statistic, showing that the independent variable, S (team stability factor), has statistical significance, and *S* is in positive correlation with the dependent variable, *Learning score*. Meanwhile, the F statistic implies that the model is acceptable. However, $R^2=0.007$, this means *S* cannot account for *Learning score* independently since there is a strong correlation between other variables and *Learning score*. In other words, *S* and *Leaning score* have a linear relation, but the goodness of fit is not satisfactory.

*Table 5. Parameter estimation and goodness of fit of the model*

|  | Coefficients | T-value | Sig |
|---|---|---|---|
| **Constant** | 82.114 | 160.756 | 0.000 |
| S | 0.502 | 2.065 | 0.039 |
| F | 4.265 | **Sig** | 0.039 |
| $R^2$ | 0.007 | Adj-$R^2$ | 0.006 |

From the regression model, we can see that value *S* has inconspicuously positive effect on learning score. This shows that *team stability* has positive influence on the academic scores of the team members, but it means little. The reasons for this phenomenon may be complex, such as personality, environment, and we will explain the details of this problem in this paper.

*4.2 Correlation of team stability with learning satisfaction*

To better test the rationality of the positive relevance between *team stability* and learning score, we delivered questionnaires to some teachers and students engaging in the consecutive course projects to find out their ideas on team stability and the possible problems in the entire learning procedure. Our questionnaires are mainly about students' opinion on learning satisfaction, so we surveyed students' perspective on the relation between team stability and learning satisfaction.

We have randomly selected 18 team leaders of the sophomores (just finishing their *database* course project), 12 team leaders of the juniors (just finishing the *combined* course project), and 15 teacher assistants (TAs) guiding course projects of the SoS@HIT who have engaged in course projects in the summer of 2011 for the questionnaire survey. Of the 45 questionnaires distributed, 45 questionnaires were returned and all questionnaires were completed and usable for data analysis, showing an effective response rate of 100 percent. Some major questions and their analysis results are shown in Fig. 2 and Fig. 3.

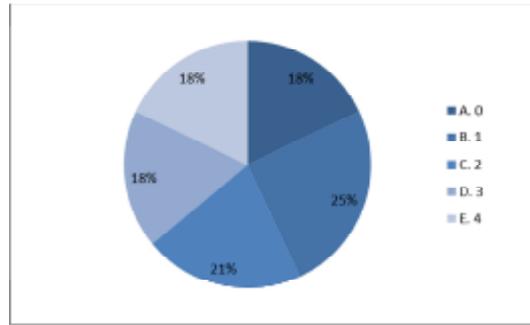

*(a) In the current course project, how many member(s) come from the prior team except for yourself?*

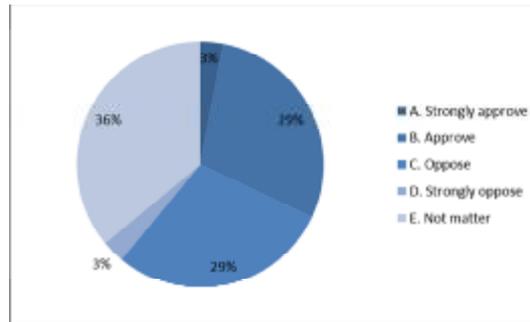

*(b) Do you approve to keep team stable in every cooperative learning activity?*

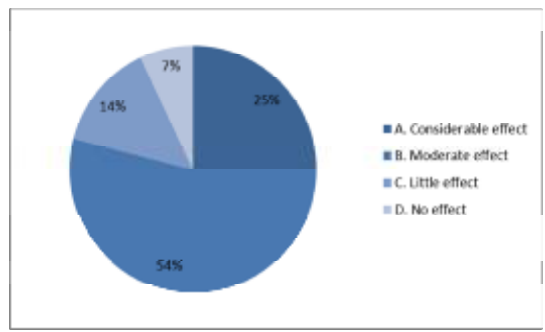

*(c) Do you think the stable teaming has an effect on higher scores?*

*Fig. 2. Some questions in questionnaire and the survey results (from students)*

The conclusion about the students can be seen in Fig. 2 as follows:

- In the practical course projects, the teaming of students has tremendous aimlessness and randomness (Fig. 2a), and most students have not realised the value of stable team (Fig. 2b)on learning satisfaction;

- Merely 25% of the students have confirmed that stable teaming can bring a positive effect on their studies (Fig. 2c);

- As for the question "whether to choose the team member in line with personal emotions (such as the friends or roommates on good terms etc) ", 14% of the students choose "frequently yes" and 64% students choose "occasionally yes", marking that most students have not developed the strong awareness of stable team;

- Additionally, from the interview we can know that even a minority of students bring forth the idea that "I hope to join in a new team in each new cooperative learning", this displays the students of this part do not like the stable team and are willing to cooperate with new team members.

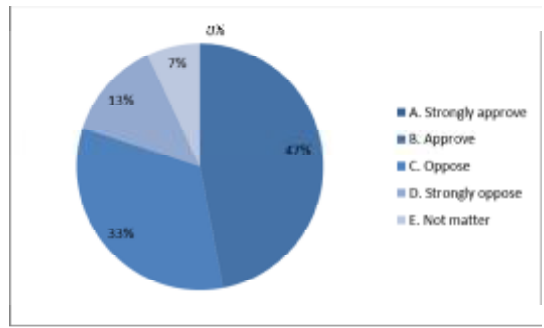

*(a) Do you approve that students should keep teams stable in every learning activity?*

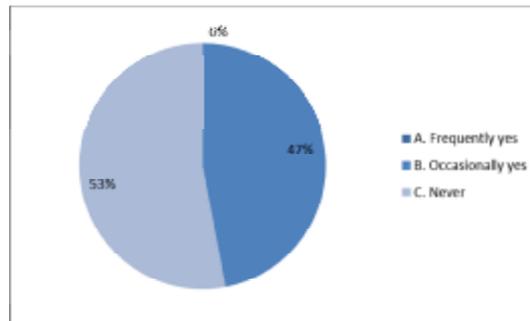

*(b) Do you often give bonus scores to the students in stable teams?*

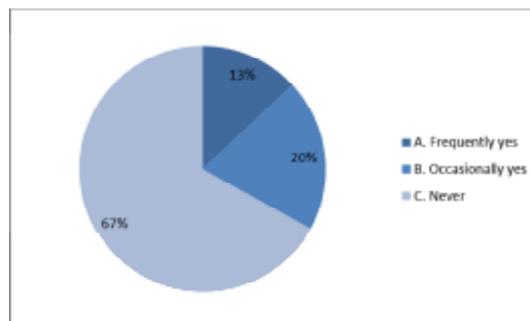

*(c) In practical course projects, do you advocate the students to maintain stable teams?*

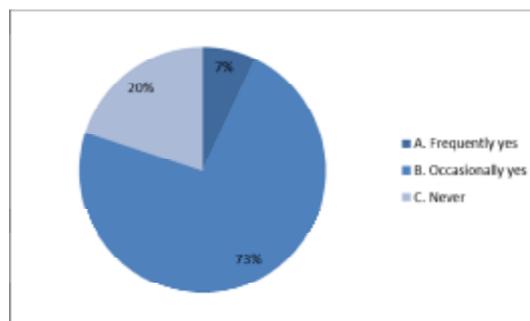

*(d) Do you give assessment to students by some subjective factors such as their performance in previous learning activities?*

Fig. 3. Some questions in questionnaire and the survey results (from teachers)

The conclusion about the teachers can be made from Fig. 3:

- Only 47% teachers hold a positive attitude toward stable teaming (Fig. 3a) and they award occasionally extra scores to a certain team for its keeping the team stability (Fig. 3b);
- Most of teachers do not advocate the students to keep stable team in practical course projects (Fig. 3c);

- Some teachers prefer to give scores to a team by its previous performance (Fig. 3d).
- The interview of the teachers says that most teachers do not prefer stable team, because they think new teams can come up numerous creative ideas. Additionally, 80% of the teachers choose "frequently yes" or "occasionally yes" about whether assessment to students by their previous performance.

Our learning satisfaction analysis shows that the attitude of participators (students and teachers) is reasonable and objective. They do not regard team stability as a factor of crucial importance but view it as a common reason that can affect learning score and improve learning satisfaction further. This attitude is consistent with our model in Formula 6, namely team stability makes inconspicuously positive effect on learning score. Therefore, most students pay little attention to keeping team stable in each cooperative learning, and many teachers never recommend students to keep stable teams or give bonus scores to the students in stable teams.

## 5. Concluding remarks and future works

Since the 1970s, cooperative learning has been promoted gradually all over the world and has been widely accepted by the educational circles of all countries. Amounts of educational and practical activities have proved its thriving vitality. This paper puts forward the approaches to measure team stability in consecutive cooperative learning. After that, it applies the practically collected data to make a basic analysis of this method and arrives at the conclusion that learning score and team stability display a weak positive correlation in practical learning. Finally, we conduct questionnaire and interview to some participants (teachers and students) in the consecutive course projects by the SoS@HIT and the result proves that major participants are satisfied with the current situation in which team stability is neither recognised well nor exactly recommended. Nevertheless, on the relation of team stability and learning performance, there are several challenging issues worthy of concerning.

Since team stability does not significantly influence on learning performance, what are the key factors that influence it significantly? In our experience, there are still some other factors influencing the learning performance. First, the leader of a team plays a crucial role in the team. He/She checks the time scheduling, controls the project quality, which decides the score of the team numbers in a great scale. Thus, the leadership of a team leader should play a critical role in learning score. Second, technical expertise is another key reason. If one team number is a high-expertise student in the software project, in technical or managerial aspect, their project will probably be excellent in quality. As a result, the final learning score will be very high. Last, teachers' impression can influence the score, too. From our interview, as listed in Fig. 3d, we know that scores are relevant to the previous impression of teachers, so we should take this factor into consideration. There into, an expected model, being interesting but challenging, is in Formula 7.

$$LP = \alpha S + \beta L + \gamma E + \lambda I \quad (7)$$

In Formula 7, $S$ means the team stability as mentioned in this study, $L$ means the leadership of team leader, $E$ marks the expertise of key member in a team, $I$ stands for the teachers' impression to a team, and $\alpha, \beta, \gamma, \lambda$ are the correlation coefficients.

Why is team stability not very important to students? Through observation and data analysis mentioned above, we recognised a surprising phenomenon that a certain number of students would rather build up a new team in succeeding activity than stay in a stable team. As mentioned in the literature, the relation of team stability with team learning and performance is a matter of some debate. On the one hand, keeping the same team members together facilitates coordination of interdependent work. On the other hand, overtime, stable teams may become slaves to routine and fail to respond to changing conditions (Edmondson, Bohmer, & Pisano, 2001). Through the interview and preliminary analysis, we gain partial reasons: i) some students try to enter a new team merely because of curiosity; ii) the minority of students want to change a role in a new project because they dislike to play the same role, such as designer, programmer, document writer or tester, in a stable team; iii) since teachers do not recommend students to keep team stable, students make a little sense of team stability and its advantage. Anyway, besides team stability, team instability is also an interesting topic worthy of concerning in future work.